\documentclass[journal,twoside,web]{color}
\usepackage{generic}
\usepackage{cite}
\usepackage{amsmath,amssymb,amsfonts}
\usepackage{bbold}
\usepackage{algorithmic}
\usepackage{graphicx}
\usepackage{textcomp}
\usepackage{steinmetz}
\usepackage{comment}
{\begin{list}{$\bullet$ \hfill}{
			\setlength{\leftmargin}{\parindent}
			\setlength{\parsep}{0.07\baselineskip}
			\setlength{\itemsep}{0.7\parsep}
			\setlength{\labelwidth}{\leftmargin}
			\setlength{\labelsep}{0em}}}{\end{list}}
\def\BibTeX{{\rm B\kern-.05em{\sc i\kern-.025em b}\kern-.08em
    T\kern-.1667em\lower.7ex\hbox{E}\kern-.125emX}}
{Soylu and Oelze
: A Data-Efficient Deep Learning Strategy for Tissue Characterization: Zone Training}
\begin{document}
\title{A Data-Efficient Deep Learning Strategy for Tissue Characterization via Quantitative Ultrasound: Zone Training}
%Velocity Filtering for Improved Resolution \\ in Ultrasound Localization Microscopy}
\author{Ufuk Soylu and Michael L. Oelze
\thanks{This work was funded by grants from the NIH (R01CA251939,R01CA273700, R21EB023403, R21EB030743)}
\thanks{Ufuk Soylu and Michael Oelze are with the Department of Electrical and Computer Engineering, and the Beckman Institute, University of Illinois at Urbana-Champaign, Urbana, IL 61801 USA. 
(e-mail: usoylu2@illinois.edu; oelze@illinois.edu).}
}
\maketitle
\begin{abstract}
Deep learning (DL) powered biomedical ultrasound imaging is an emerging research field where researchers adapt the image analysis capabilities of DL algorithms to biomedical ultrasound imaging settings. A major roadblock to wider adoption of DL powered biomedical ultrasound imaging is that acquisition of large and diverse data-sets is expensive in clinical settings, which is a requirement for successful DL implementation. Hence, there is a constant need for developing data-efficient DL techniques to turn DL powered biomedical ultrasound imaging into reality. In this work, we develop a data-efficient deep learning training strategy for classifying tissues based on the ultrasonic backscattered RF data, i.e., quantitative ultrasound (QUS), which we named \textit{Zone Training}. In \textit{Zone Training}, we propose to divide the complete field of view of an ultrasound image into multiple zones associated with different regions of a diffraction pattern and then, train separate DL networks for each zone. The main advantage of \textit{Zone Training} is that it requires less training data to achieve high accuracy. In this work, three different tissue-mimicking phantoms were classified by a DL network. The results demonstrated that \textit{Zone Training} can require a factor of 2-3 less training data in low data regime to achieve similar classification accuracies compared to a conventional training strategy.   
\end{abstract}

\section{Introduction}
\label{sec:intro}
Deep learning (DL) powered biomedical ultrasound imaging is becoming more advanced and coming closer to routine clinical applications in recent years \cite{akkus2019survey}. DL is the process of learning a hierarchy of parameterized nonlinear transformations to perform a desired function. Therefore, DL algorithms extract a hierarchy of features from raw input images and image data automatically rather than extracting features manually. Due to rapid increase in computational power and large data-sets, DL and machine learning algorithms have emerged as leading tools and have achieved impressive results in various research fields. Among DL algorithms, convolutional neural networks (CNN) use convolutional layers to embed structural priors of translational invariance, which make them parameter and data efficient learners for image analysis tasks. Respectively, CNNs are the most popular and successful DL structure for ultrasound biomedical imaging \cite{liu2019deep}. Common DL applications that have provided notable results in the context of biomedical ultrasound imaging are classification\cite{liu2019deep, liu2015tumor, shi2016stacked, nguyen2021use, han2020noninvasive, byra2022liver}, detection\cite{liu2019deep, cao2017breast}, segmentation\cite{liu2019deep, carneiro2011segmentation, carneiro2013combining, zhang2016coarse, baka2017ultrasound} and image reconstruction \cite{van2019deep, vedula2018learning,perdios2017deep,simson2019end,khan2019universal,luchies2018deep,hyun2019beamforming}. Furthermore, DL algorithms have been employed in advanced ultrasound imaging applications such as super-resolution imaging of microvasculature structure via Ultrasound Localization Microscopy \cite{liu2020deep, van2020super}.

Even though DL is promising for biomedical ultrasound imaging, there are certain roadblocks to wider adoption. A major roadblock is that acquiring large and diverse data-sets is expensive. Hence, it is important to develop data-efficient DL algorithms to overcome this limitation. Another roadblock is that there are large variations in ultrasound images due to operator, patient or machine dependent factors. Therefore, improving the robustness of DL algorithms against variations in ultrasound images is necessary. Overall, to turn DL powered biomedical ultrasound imaging into reality, there is a constant need for developing DL algorithms, which are data efficient and more robust against variations in ultrasound images. 

In this paper, we examine DL techniques for classifying samples based on ultrasonic backscattered RF data similar to the work of Nguyen et al. \cite{nguyen2021use}. Classifying tissues has recently evolved from model-based approaches such as quantitative ultrasound (QUS) techniques to model-free, DL-based techniques. Nguyen et al. \cite{nguyen2019characterizing} demonstrated that QUS techniques are able to detect the presence of steatosis in a rabbit model of fatty liver with a classification accuracy of 84.11$\%$. In a later study, Nguyen et al. \cite{nguyen2021use} compared a DL-based classifier to a QUS-based classifier for the problem of fatty liver classifier and found that the DL-based classifier outperformed the QUS-based approach with the accuracy of 74$\%$ versus 59$\%$. While the traditional spectral-based QUS approach does not utilize the phase information in the RF signal, DL-based approaches can extract additional classification power from the lost phase information from the RF data. Furthermore, the DL-based approach does not require a model like the QUS approach, which means that features of the backscattered signal that are missed by the QUS approach can be picked up by the DL approach. Subsequently, the DL approach performs feature extraction and classification simultaneously.

To improve classification, we consider the diffraction patterns associated with ultrasonic transducers and how they result in different regions or 'zones' that must also be learned to separate the system signal from the sample signal. We propose a training strategy, which we call \textit{Zone Training}. In \textit{Zone Training}, we propose to divide the complete field of view of an ultrasound image into multiple zones such as pre-focal, on focus and post focal zones. Then, we train separate neural networks for each zone by using the data belonging to the corresponding zone. In a sense, we train expert neural networks for each zone as opposed to \textit{Regular Training}, which uses all data coming from the complete field of view to train a single neural network. The main intuition is that at each zone, there are different diffraction patterns and learning all the patterns by a single network is harder than learning a single diffraction pattern by a single expert network. The main advantage of \textit{Zone Training} is that it requires less data to achieve similar classification performance in comparison to \textit{Regular Training} in low data regime.

\section{Background and Motivation}
\label{sec:background}

\textit{Zone Training} is similar to applying an attention mask to the input manually and training separate networks for each mask to learn dedicated convolutional filters per zone. In this sense, \textit{Zone Training} applies attention in a simple and direct way to incorporate the physics of diffraction into DL training. There are methods in the literature that enable learning of varying convolution kernels over the complete field of view, e.g., pixel-adaptive convolution \cite{su2019pixel}. In general, attention mechanisms, e.g. self-attention mechanisms,\cite{guo2022attention} were invented initially for computer vision tasks, where the data is abundant and they apply attention by altering network architecture and hence model complexity to improve classification accuracy. However, in the context of biomedical imaging, we are in a different regime where the data is often scarce. Therefore, we favor utilizing a smaller training set to achieve a desired classification accuracy. Overall, \textit{Zone Training} provides us a method to reduce training set size by modifying data distribution without altering model complexity. Furthermore, \textit{Zone Training} can be perceived as utilizing a symbolistic approach, in the form of a simple if-else structure (if data is from a certain zone, train a specific network), to transfer physics knowledge into DL training. Combining DL and symbolic reasoning is known in the literature as Neural-Symbolic Computing (NSC), which can lead to data-efficient AI \cite{adadi2021survey}.

\begin{figure}[hbt!]
\begingroup 
    \centering
    \begin{tabular}{cc}
\hspace{-0.5cm}{ \includegraphics[width=0.88\linewidth]{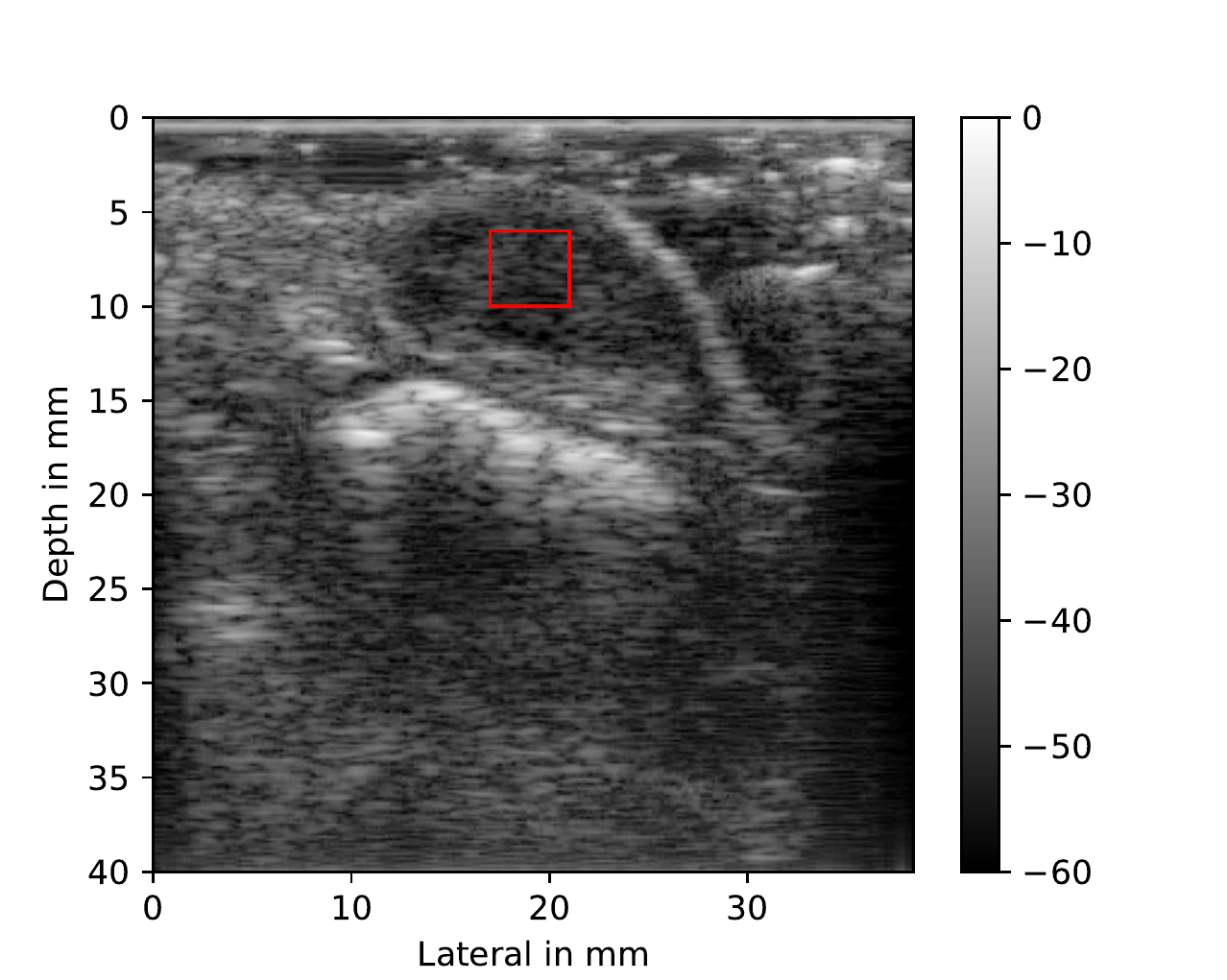}} \\ 
\hspace{-0.5cm}{ \includegraphics[width=0.88\linewidth]{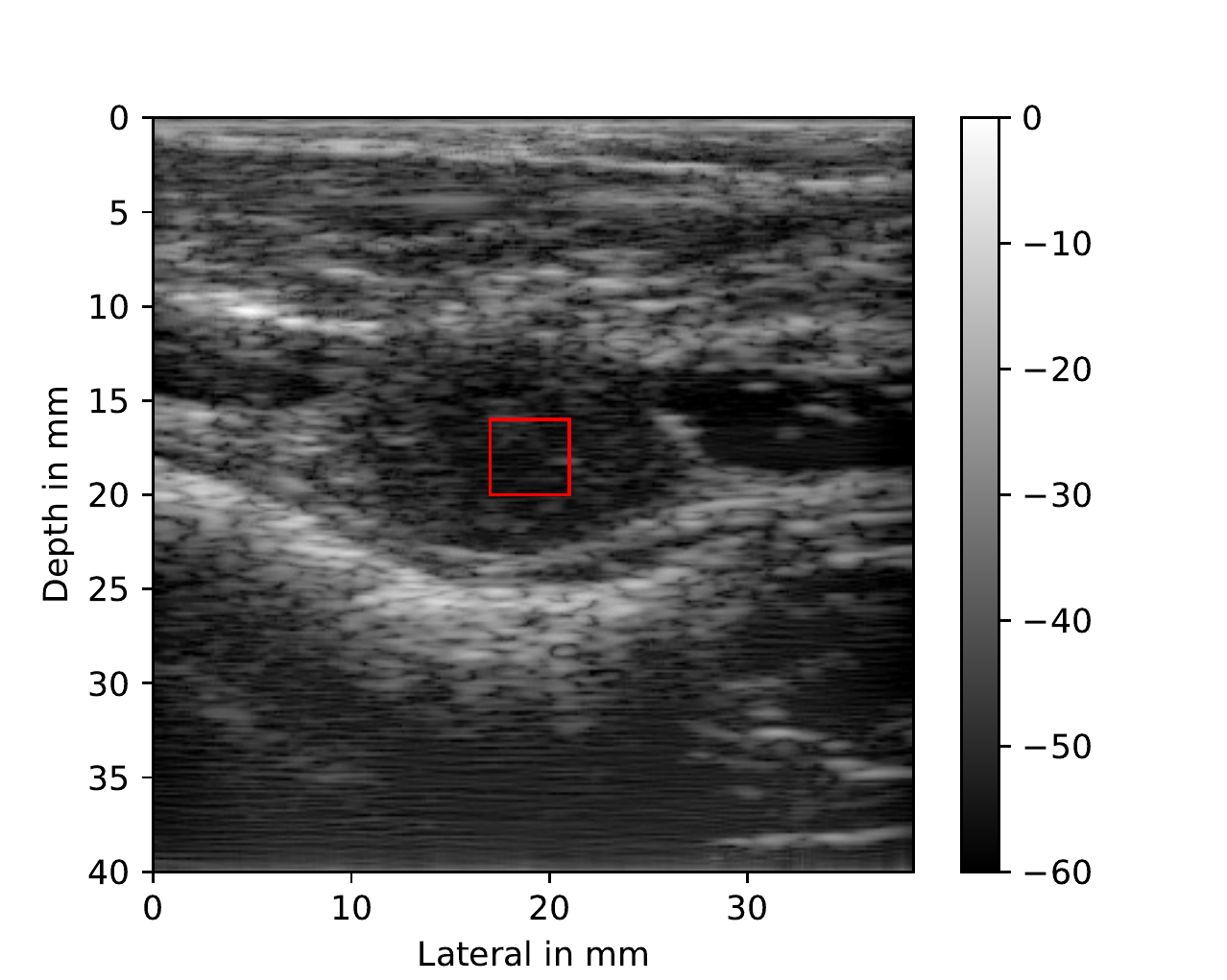}}\\
\end{tabular}
\caption{Tumor examples, in dB scale, whose scan depths vary in the field of a transducer probe. The tumor images were acquired from rabbits having mammary VX2 tumors. All animal experiments were approved by the Institutional Animal Care and Use Committee at the University of Illinois at Urbana–Champaign.}
\label{fig:tumor}
\endgroup
\end{figure}

In our experiments, we chose tissue classification as the primary application and tested our proposed method to classify three distinct tissue-mimicking phantoms. To further motivate \textit{Zone Training}, we describe a clinical scenario when it is the most relevant and advantageous. For instance, ultrasound imaging can be used to examine and characterize tumors, whether benign or malignant, which can exist at different depths within a body. When using QUS approaches for tumor characterization, a region of interest (ROI) inside the tumor is selected to examine the signals from the tumor. We show two tumor image examples where the tumors are at different depths in Fig. \ref{fig:tumor}, but the same probe is used. Different depths correspond to different zones and red rectangles are sampled from the tumors in those ultrasound images. In \textit{Zone Training}, we have separately trained DL algorithms for each zone. Experimental studies for \textit{Zone Training} in this work were conducted under two assumptions following the clinical scenario. Firstly, we assume that we are not trying to detect the ROI. In other words, we are given rectangular patches of data to classify a tissue state. The ROI can be detected by another algorithm or by the operator. The operator in the clinic can adjust imaging settings to obtain the best imaging quality, and then, select the ROI which should be considered as Human-centered AI, whose aim is to amplify and augment rather than displace human abilities \cite{shneiderman2021human}. The second assumption is that the ROI is larger than the rectangular patches of data so that the classification networks in this work, take uniform rectangular patches as their input. 

\section{Methods}
\label{sec:met}
\subsection{Phantoms}
\begin{figure}[hbt!]
\begingroup
    \centering
    \begin{tabular}{c c c}
\hspace{-0.5cm}{ \includegraphics[scale=0.14]{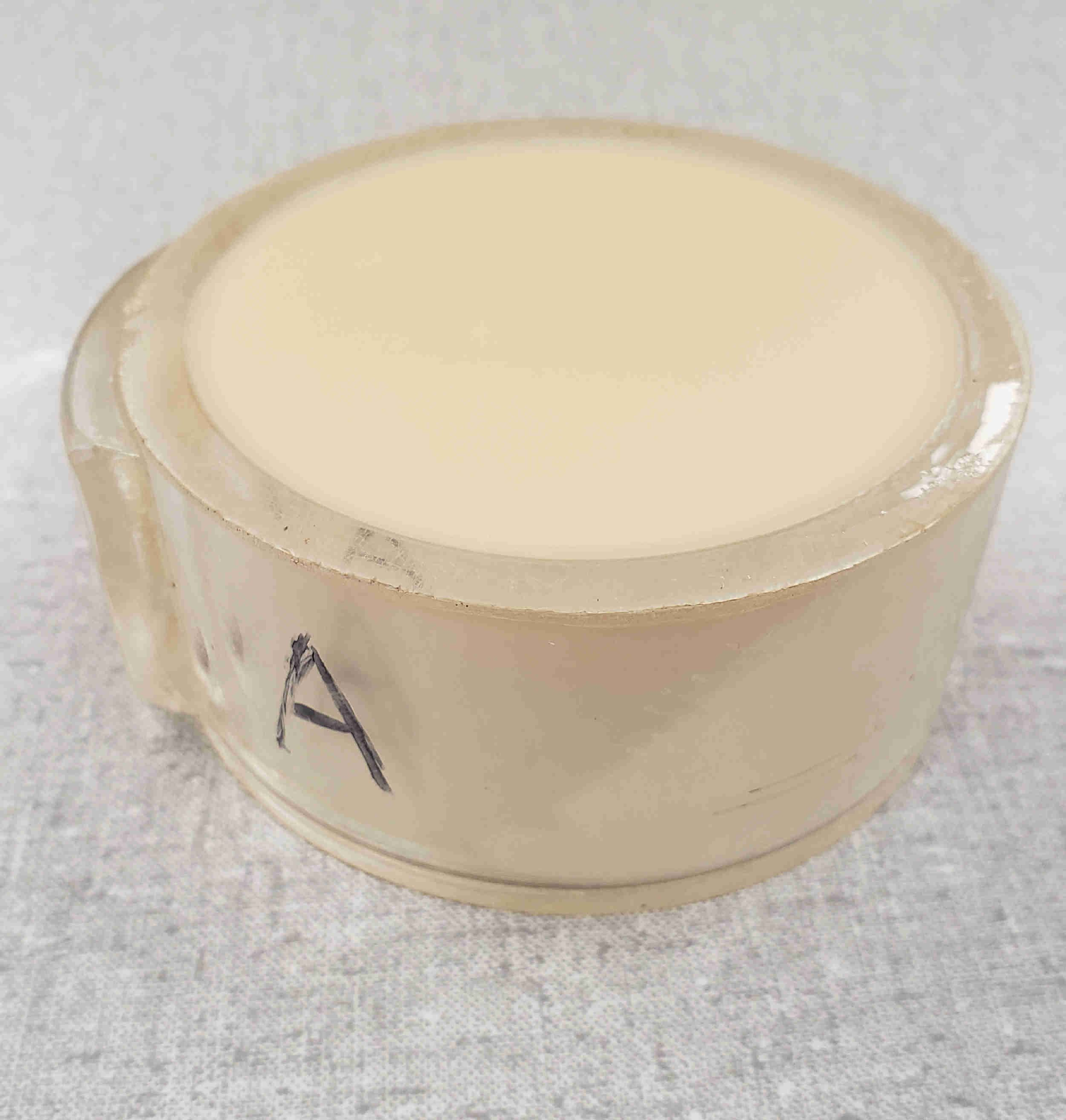}}&\hspace{-0.5cm}{ \includegraphics[scale=0.134]{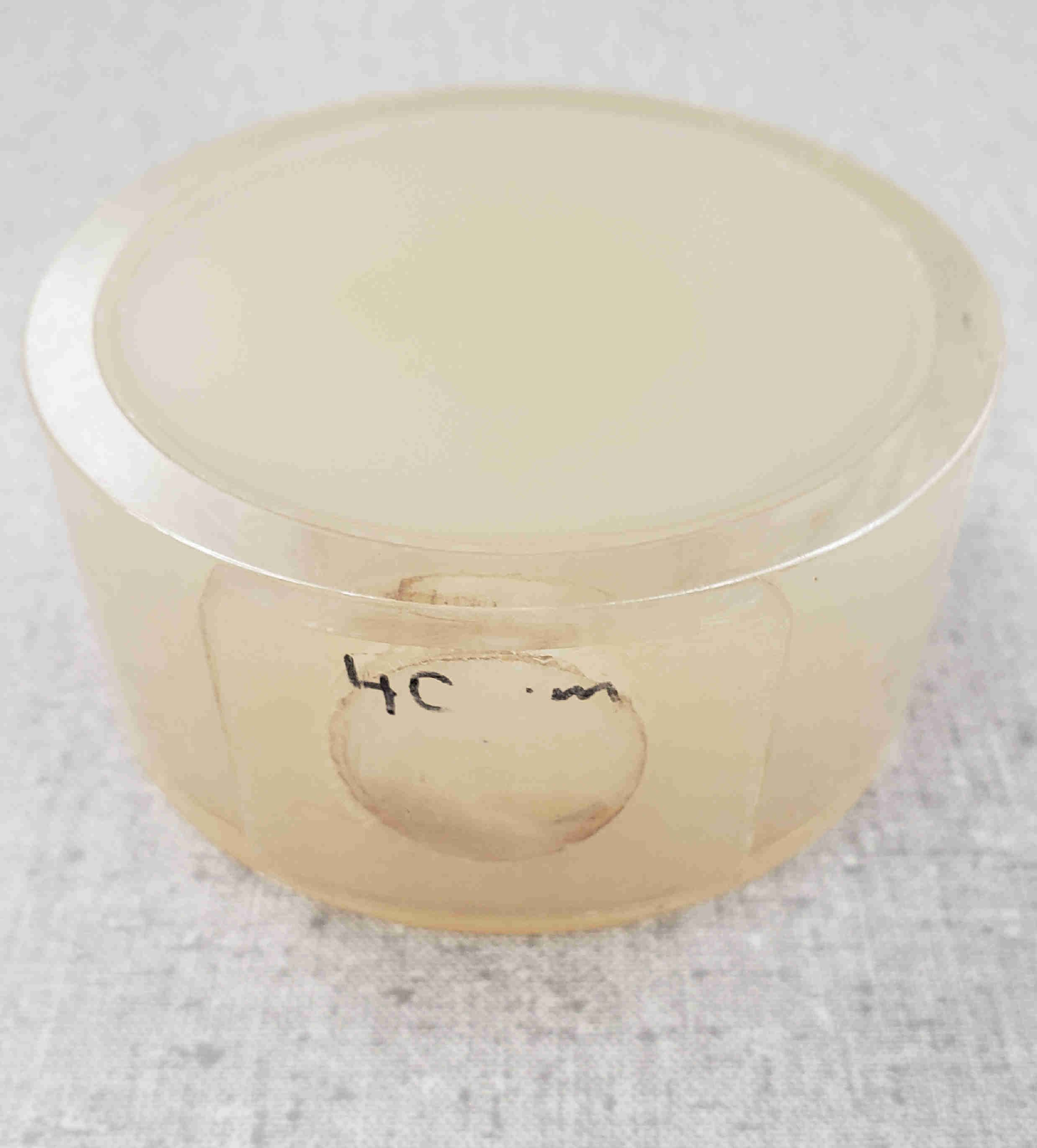}} &\hspace{-0.5cm}{ \includegraphics[scale=0.130]{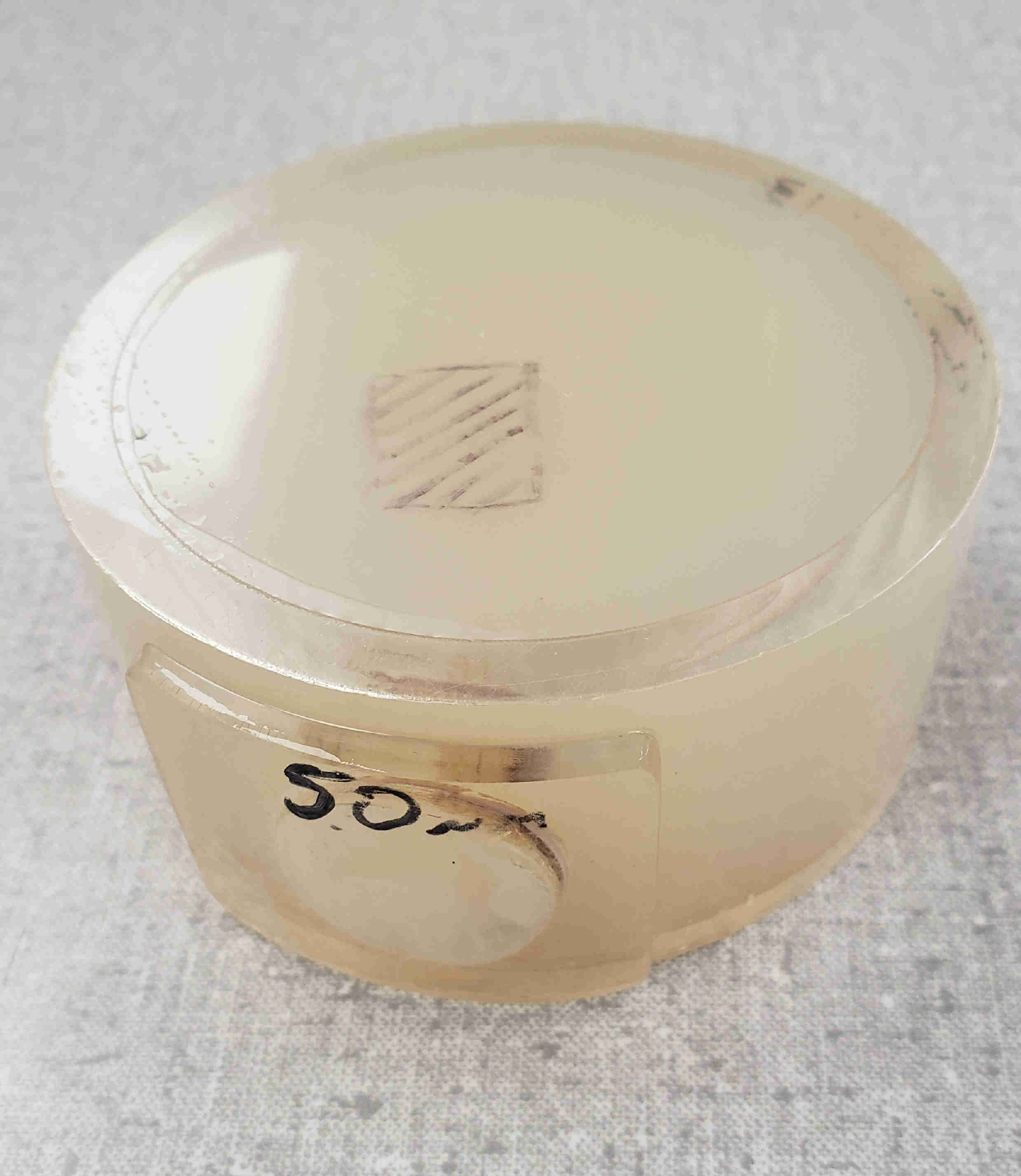}}\\
\hspace{-0.5cm} Phantom1 &\hspace{-0.5cm} Phantom2 &\hspace{-0.5cm} Phantom3
\end{tabular}
\caption{Tissue-mimicking Phantoms}
\label{fig:phantom_picutes}
\endgroup
\end{figure}

Three different tissue-mimicking phantoms were used in the experiments, which we designated as Phantom1, Phantom2 and Phantom3. They are cylindrically shaped as shown in Fig. \ref{fig:phantom_picutes} and their properties are summarized in Table \ref{table:phantom_properties}. 

Phantom1, which mimics human liver, has been described by Wear et al. \cite{wear2005interlaboratory}. Phantom1 had a measured attenuation coefficient slope of approximately 0.4 dB$\times$cm$^{-1}\times$MHz $^{-1}$. 
%with a scatterer number density of approximately 5 scatterers/mm$^3$. 
Its materials were produced based on the method of Madsen et al. \cite{madsen1998liquid} and they are macroscopically uniform. The only nonuniformity in Phantom1 results from the random positioning of microscopic glass bead scatterers. The component materials and their relative amounts by weight for Phantom1 are agarose (3.5$\%$), n-propanol (3.4$\%$), 75 to 90 $\mu$m-diameter glass beads (0.38$\%$), bovine milk concentrated 3 times by reverse osmosis (24.5$\%$), liquid Germall Plus preservative (International Specialty Products, Wayne, NJ) (1.88$\%$), and 18-$M\Omega$-cm deionized water (66.3$\%$)."

Phantom2 and Phantom3 are both low attenuation phantoms, whose properties have been described by Anderson et al. \cite{anderson2010interlaboratory} and constructions have been described Madsen et al. \cite{madsen1978tissue}. Both phantoms were made with the same
weakly-scattering agar background material but contained different sizes of scatterers. They have an attenuation coefficient slope of approximately equal to 0.1 dB$\times$cm$^{-1}\times$MHz $^{-1}$. %Additionally, their scatterer number densities are about 2.23 g/L. 
Glass-sphere scatterers (Potters Industries, Inc., Valley Forge, PA; Thermo Fisher Scientific (formerly Duke Scientific), Inc., Waltham, MA) were used in both phantoms with weakly scattering 2$\%$ agar background. The only difference in the phantoms was the size distribution of the glass bead scatterers, i.e., Phantom2 had a mean diameter of 41 $\mu$m and Phantom3 had a mean diameter of 50 $\mu$m. 

\begin{table}[htb!]
    \centering
\caption{Phantom Properties}
{\small
\begin{tabular}{ |p{2.9cm}||p{1.6cm}|p{1.2cm}|p{1.2cm}|}
 \hline
& Phantom1 & Phantom2 & Phantom3\\
\hline
%Number density  & 5beads/mm$^3$ & 2.23g/L& ?? \\
%\hline
Sphere diameter ($\mu $m) & 75-90 & 41 $\pm$ 2 & 50 $\pm$ 2.4\\
\hline
Background material & 3.5$\%$ agar & 2$\%$ agar& 2$\%$ agar \\
\hline
Sound speed (m/s) & 1540  & 1539&  1539\\
\hline
Attenuation (dB/cm/MHz)& 0.4 & 0.1& 0.1 \\
\hline
\end{tabular}
}
\label{table:phantom_properties}
\end{table}

\subsection{Ultrasound Scanning Procedures}

Ultrasound gel was placed on the surfaces of the phantoms and then the phantoms were scanned with an L9-4/38 transducer using a SonixOne system (Analogical Corporation, Boston, MA, USA) providing an analysis bandwidth of 2-7.5 MHz. 1,007 frames of post-beamformed RF data sampled at 40 MHz were acquired from each phantom and saved for offline processing.

The imaging array had a center frequency of approximately 5.5 MHz and was operated with a single axial focus at 2 cm depth and a fixed elevational focus of 1.9 cm. The center frequency of the pulse was chosen as 9 MHz to provide higher bandwidth (resolution) for the transducer. The total imaging depth was chosen as 4 cm, which is equal to the height of the phantoms. Output power was chosen as -5 dB, which corresponds to -5 dB lower power level with respect to maximum output power of the system. 

% \begin{table}[htb!]
%     \centering
% \caption{Imaging Settings for SonixOne System }
% {\small
% \begin{tabular}{ |p{3.5cm}||p{3cm}|}
%  \hline
%  Imaging Parameters& Values \\
%  \hline
% Center Frequency of Pulse& 9 MHz\\
% Pulse Shape & Cosine signal \\
% Pulse Duration & 1 Period \\
% Focus & Single at 2 cm\\
% Output Power & -5 dB\\
% Imaging Depth & 4 cm\\
%  \hline
% \end{tabular}
% }
% \label{tab:imaging_settings}
% \end{table}

\subsection{Data-set}
\label{section:dataset}

We acquired 1,007 ultrasound images per phantom by free-hand motion. In total, we acquired 3,021 ultrasound frames. The size of an ultrasound image frame was 2,080 pixels$\times$256 pixels. There were 2,080 samples along the axial direction that corresponded to 4 cm depth. Even though the L9-4/38 transducer has 128 channels, the SonixOne system interpolates to 256 channels that  correspond to 256 lateral pixels. The data-set of ultrasound images is also publicly available at https://osf.io/7ztg3/ (DOI 10.17605/OSF.IO/7ZTG3). After acquiring the ultrasound images per phantom, we extracted rectangular image patches to be used in training and testing.

\begin{figure}[hbt!]
\begingroup
    \centering
    \begin{tabular}{cc}
\hspace{-0.95cm}{ \includegraphics[width=0.68\linewidth]{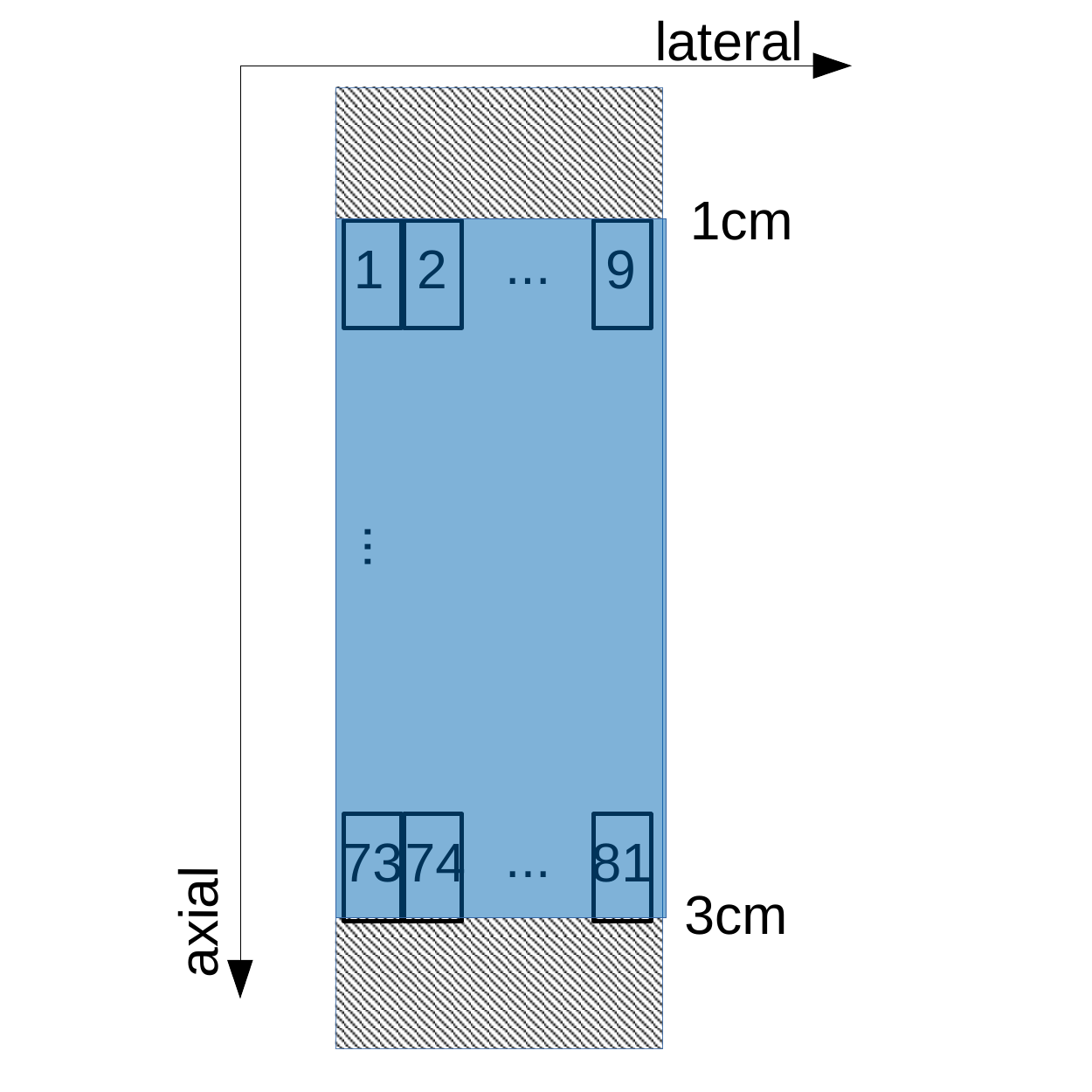}}& \hspace{-2.0cm}{ \includegraphics[width=0.68\linewidth]{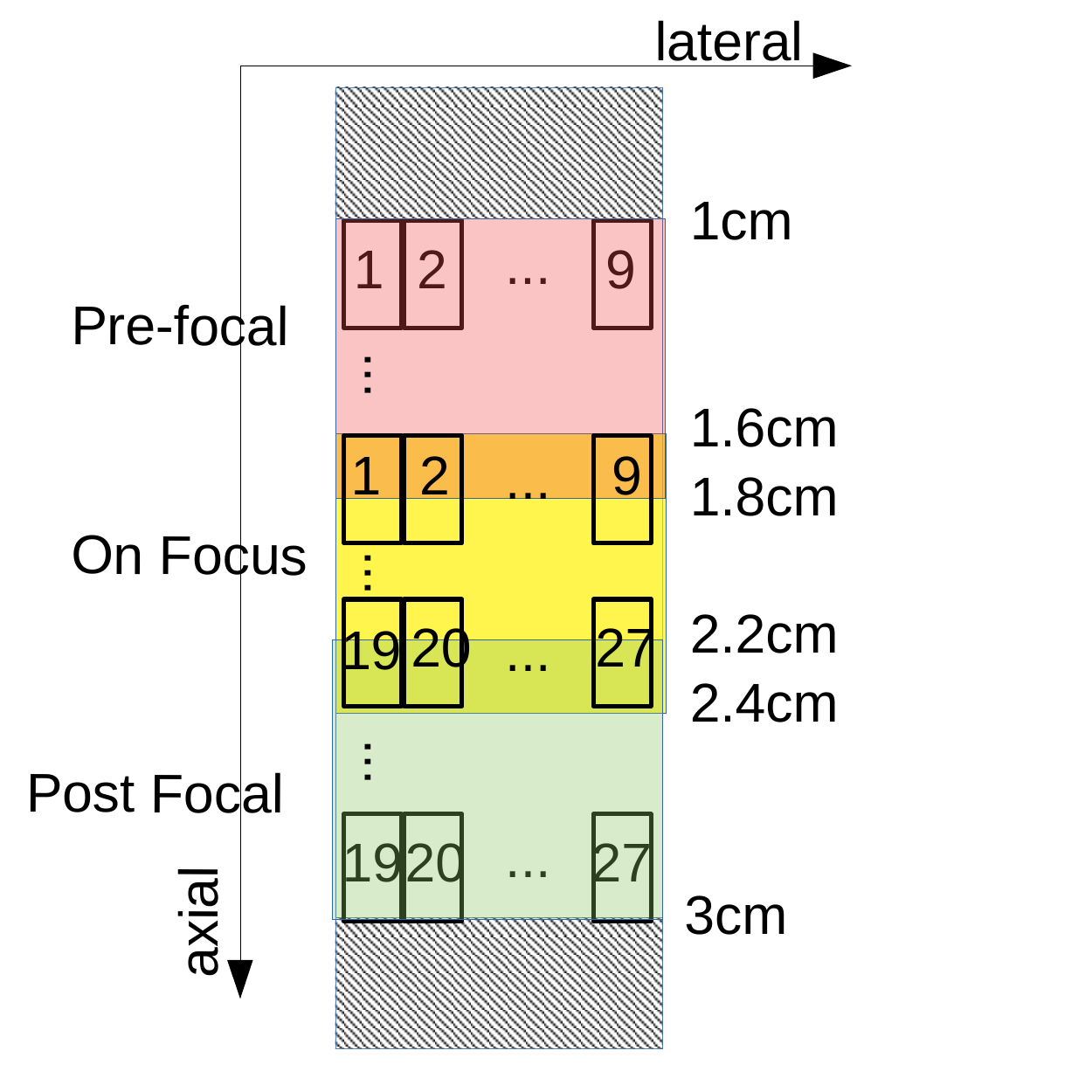}} \\
\hspace{-1.2cm} Regular Training & \hspace{-1.5cm} Zone Training 
\end{tabular}
\caption{Patch Extraction}
\label{fig:horizontal_patch_extract}
\endgroup
\end{figure}

In patch extraction, which is depicted in Fig. \ref{fig:horizontal_patch_extract}, we extracted rectangular image patches whose sizes were 200 pixels$\times$26 pixels that correspond to square image patches whose size were 4 mm$\times$4 mm in physical dimensions. From one ultrasound image, we could extract 81 (9 lateral $\times$ 9 axial) image patches when we used the complete field of view as in \textit{Regular Training}. While extracting image patches, we didn't use the first 540 pixels in the ultrasound image. Axially, we obtained the next line of individual patches by translating the start of the next patch by 100 pixels along the axial depth. Laterally, we obtained the next line of individual patches by  translating the start of the next patch by 26 pixels along the axial depth. Overall, in patch extraction for \textit{Regular Training}, there was 9 axial lines and 9 lateral lines to extract individual patches that lead to extracting 81 image patches per ultrasound image.

\begin{figure}[hbt!]
\begingroup
    \centering
    \begin{tabular}{c}
\hspace{-0cm}{ \includegraphics[width=0.7\linewidth]{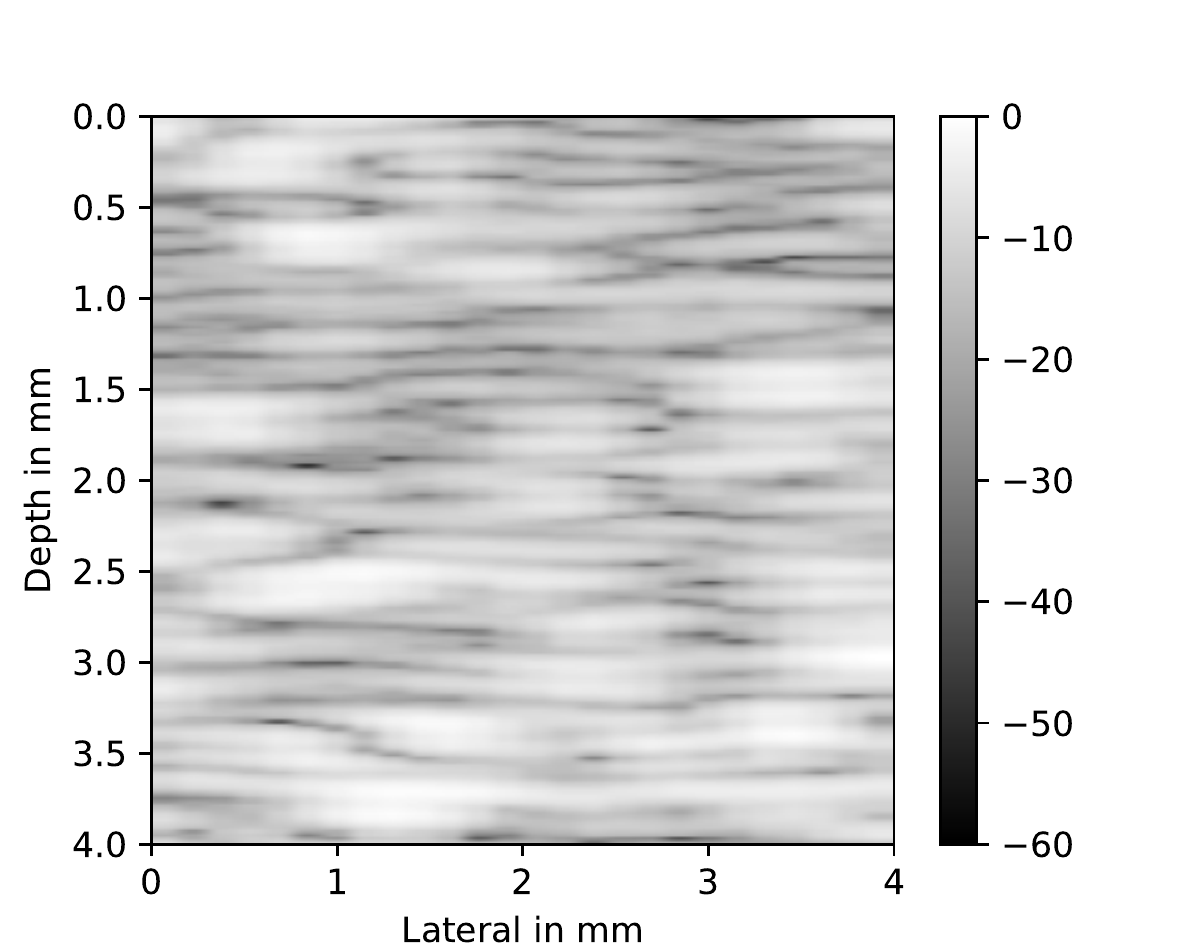}}\\ \hspace{-0cm}{ \includegraphics[width=0.7\linewidth]{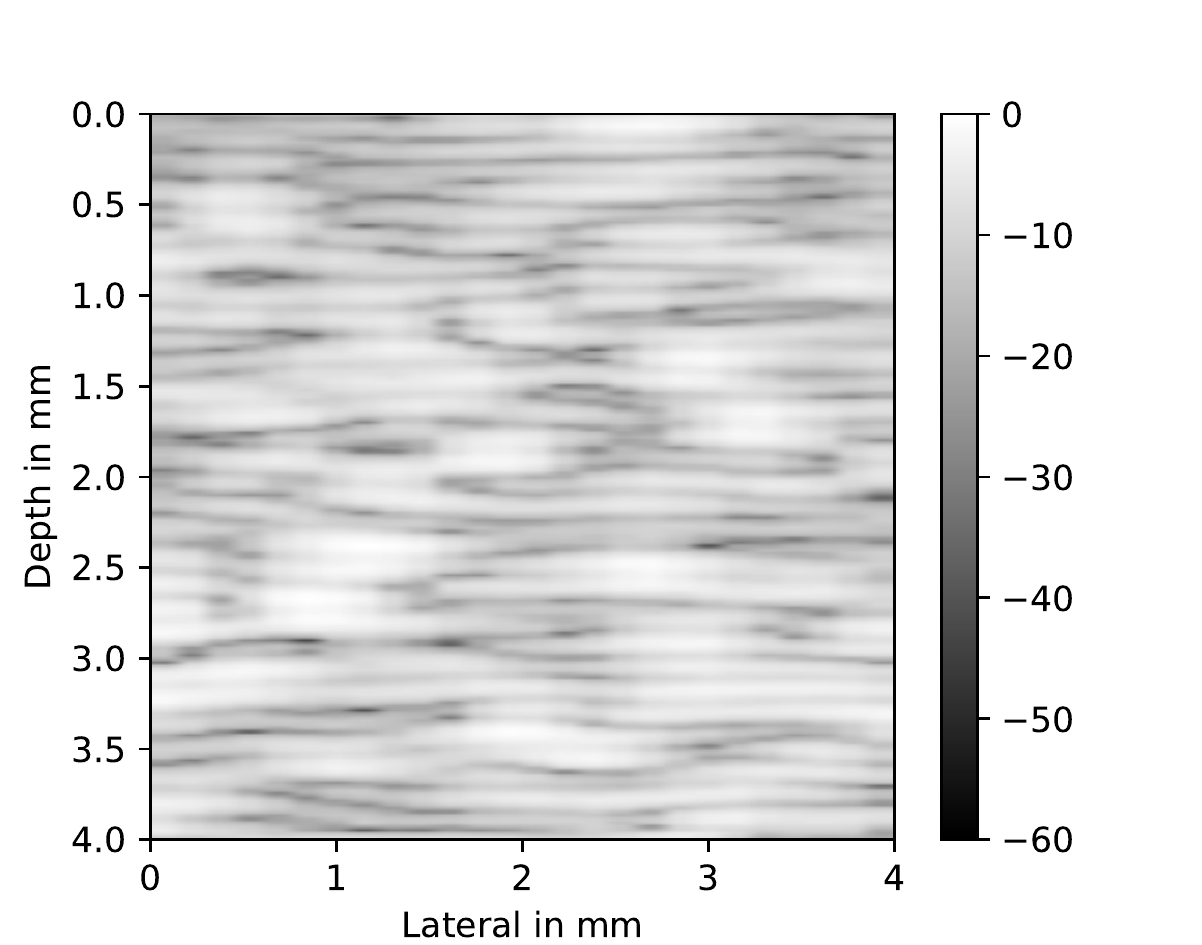}}\\ \hspace{-0cm}{ \includegraphics[width=0.7\linewidth]{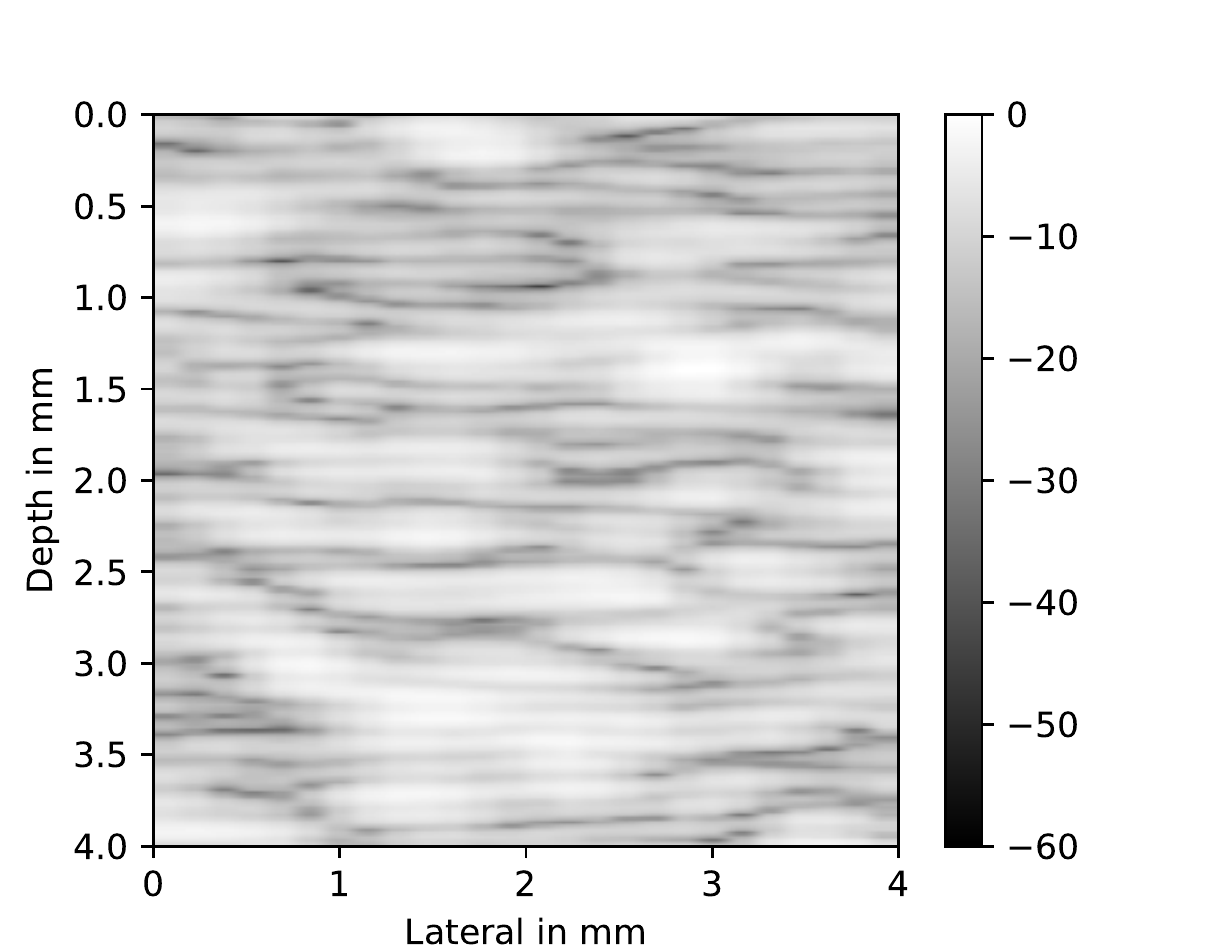}}
\end{tabular}
\caption{Example B-mode images of extracted patches in dB scale from the pre-focal zone centered at 1.4 cm. Classifying these patches are difficult by visual inspection. Top Row: Phantom1; Middle Row: Phantom2 and Bottom Row: Phantom3.}
\label{fig:dataset_examples}
\endgroup
\end{figure}

For \textit{Zone Training}, we first developed definitions for the zones based on the diffraction pattern for a single focused transducer. In this work, we broke the complete field of view into three zones axially: a pre-focal zone which is centered at 1.4 cm, an on focus zone which is centered at 2 cm, a post focal zone which is centered at 2.6 cm. Then, each zone coincides with 3 axial lines of the complete field of view in patch extraction. Therefore, three zones together use the same data as in \textit{Regular Training}. For each zone, we extracted 27 (9 lateral $\times$ 3 axial) image patches whose sizes were 200 pixels$\times$26 pixels corresponding to 4 mm$\times$4 mm in physical dimensions. Example B-mode images of the patches corresponding to each phantom are provided in Fig.\ref{fig:dataset_examples}.

\subsection{Training}

DL training was done by using two machines each with a single GPU. One machine had TITAN RTX and the other machine had RTX A5000. All implementations were done with the PyTorch library \cite{paszke2019pytorch}. As a data preprocessing step, we applied z-score normalization at the patch level, i.e., the mean intensity value of patches was subtracted from each patch, and then, each pixel in a patch was divided by the standard deviation of the intensity of the patches. The batch number was chosen as 128 through out all experiments. Horizontal flip with 0.5 probability was implemented as a data augmentation step in the training process. We used the Adam algorithm \cite{kingma2014adam} as the optimizer in all experiments. Additionally, the models were trained by using cross entropy loss with uniform class weights, which includes built-in softmax function in PyTorch implementation \cite{paszke2019pytorch}.
% \begin{equation}
% \label{eq:cross}
%     l(x,y) = \sum_{n=1}^{N} \frac{-log \frac{exp(x_{n, y_n})}{\sum_{c=1}^{C} exp(x_{n,c}) }}{N}
% \end{equation}
% where $C$ is the number of classes, $N$ is the batch size, $x_{n,c}$ is the logit, which corresponds to class $c$ and $x_{n,y_n}$ is the logit with correct class $y_n$.

In this work, we used CNN architectures consisting of two parts: feature extractors that consist of convolution layers, max-pooling layers and non-linear activation functions, and a classifier that consists of fully connected layers and non-linear activation functions. They also have significantly fewer parameters and so they can be trained more efficiently than fully connected networks \cite{alom2018history}. We used a slightly modified CNN architecture, which is derived from AlexNet \cite{krizhevsky2012imagenet} and is shown in Table \ref{tab:network}. In the training, dropout layers with 0.5 probabilities were added to improve the regularization and deal with over-fitting, before fully connected1 and fully connected2 layers. Initial weights for the network were chosen based on the original paper \cite{krizhevsky2012imagenet}.

\begin{table}[htb!]
{\normalsize
    \centering
\caption{Network Architecture for Horizontal Patch Extraction}{$
\begin{array}{ccc}
\hline \text {Layer Name} & \text {Output Size} & \text {Regular \& Zone Training} \\
\hline \text {conv1\&relu} & 48 \times 4 \times 96 & 11 \times 11, \text {stride} 4\\
\hline \text {conv2\&relu} & 48 \times 4 \times 256 & 5 \times 5, \text {pad} 2 \\
\hline \text {conv3\&relu} & 48 \times 4  \times 384 & 3 \times 3, \text {pad} 1  \\
\hline \text {conv4\&relu} & 48 \times 4 \times 384 & 3 \times 3, \text {pad} 1  \\
\hline \text {conv5\&relu} &  48 \times 4 \times 256 & 3 \times 3, \text {pad} 1  \\
\hline \text {maxpool1} &  23 \times 1 \times 256 & 3 \times 3, \text {stride} 2 \\
\hline \text {fc1\&relu} & 4096 & 5888 \times 4096 \text { connections}\\
\hline \text {fc2\&relu} & 4096 & 4096 \times 4096 \text { connections}\\
\hline \text {fc3} & 3 & 4096 \times 3 \text { connections}\\
\hline
\end{array}$}
\label{tab:network}}
\end{table}

In the experiments, we searched the learning rate and the epoch number by using a validation set. More specifically, the learning rate and the epoch number were determined to achieve “asymptotic test accuracy”, which ideally is defined as the number of epochs of training required such that any further training provides no improvement in test accuracy. The process of forming training, testing and validation sets started with randomly selecting the desired number of ultrasound images per phantom. The same number of ultrasound frames were set apart for validation, training and testing sets. Then, we extracted patches, as described in Section \ref{section:dataset}, to form the training, testing and validation sets. After adjusting the learning rate and epoch number by using the validation set, we trained neural networks in the training sets and obtained classification accuracies in the test sets. We repeated each experiment 10 times starting from random ultrasound frame selection for training and testing sets. In Section \ref{sec:results}, we report the learning rate, epoch number, mean classification accuracy and standard deviation for each experiment. 

% As we described in Section \ref{sec:intro}, we propose to divide the complete field of view into multiple zones. Then, we train separate neural networks per each zone by using the data belonging to the corresponding zone. We name this strategy as \textit{"Zone Training"} as opposed to \textit{"Regular Training"}, which uses all data coming from the complete field of view to train a single neural network. Briefly, the main hypothesis is that at each zone, there are different diffraction patterns and learning all the patterns by a single network is harder than learning a single diffraction pattern by a single expert network.

\subsection{Depth-Aware Training}

\begin{figure}[hbt!]
\begingroup
    \centering
    \begin{tabular}{cc}
\hspace{-0.5cm}{ \includegraphics[width=0.9\linewidth]{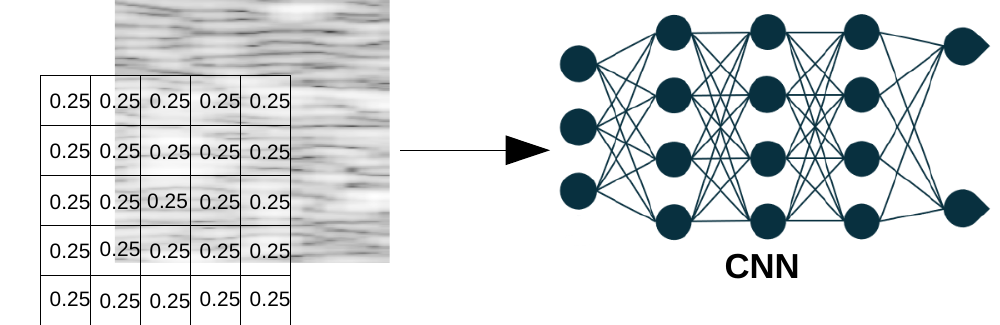}} \\
\end{tabular}
\caption{\textit{Depth Aware Training}: Two-Layered Input, one with depth information and the other one with ultrasound RF data}
\label{fig:depth_aware}
\endgroup
\end{figure}

In patch extraction, global coordinates are lost. Therefore, in addition to \textit{Regular Training}, \textit{Zone Training} is also compared against \textit{Depth Aware Training}, which utilizes global coordinates in the training. In \textit{Depth Aware Training}, we input the depth as an additional feature. Specifically, the CNN now takes a two-layered input, one layer is the image patch of 200 pixels $\times$26 pixels and the other one is a constant array of 200 pixels$\times$26 pixels whose values correspond to the relative depth, as shown simplistically in Fig. \ref{fig:depth_aware}. The depth information is normalized between 0-1 where 0 is the depth of the nearest patch and 1 is the depth of the farthest patch. Overall, \textit{Depth Aware Training} is designed to consider the global location of the input patch during both training and testing so that the DL network adapts itself based on the relative depth being near 0 or 1.

\section{Results}
\label{sec:results}

Results are organized into two parts. In the first part, we present results that help us to determine if our zone definitions are favorable by experimenting with axial zone widths, axial zone locations and by sweeping testing zone centers around training zone centers. Our purpose in this part is to determine a reasonable way to divide the field of view into multiple zones, which is required for \textit{Zone Training}. In the second part, we investigate the relationship between training set size and classification accuracy for \textit{Zone Training}, \textit{Regular Training} and \textit{Depth Aware Training}.

\subsection{Examination of Zone Definitions }
\label{sec:how_to_define}

We now present four results that are helpful in determining zone definitions. In the first result, we investigate how much classification accuracy drops as we shift the testing zone away from the training zone. Specifically, we train a neural network by using patches from the on focus zone, and then, we test the neural network with patches from nearby zones. This result shows us how much the diffraction patterns change around the focal zone. In the second result, we repeat the same experiment for the pre-focal zone and the post focal zone to investigate how much the diffraction patterns change around these zones. In the third result, we experiment with axial zone width in terms of number of overlapping patches per zone. We plot classification accuracy for the on focus zone when we increase the number of overlapping patches used in patch extraction. In the fourth result, we experiment with axial zone locations and we plot classification accuracy at different zone centers.

\begin{figure}[hbt!]
\begingroup
    \centering
    \begin{tabular}{c}
\hspace{-0.3cm}{ \includegraphics[width=1\linewidth]{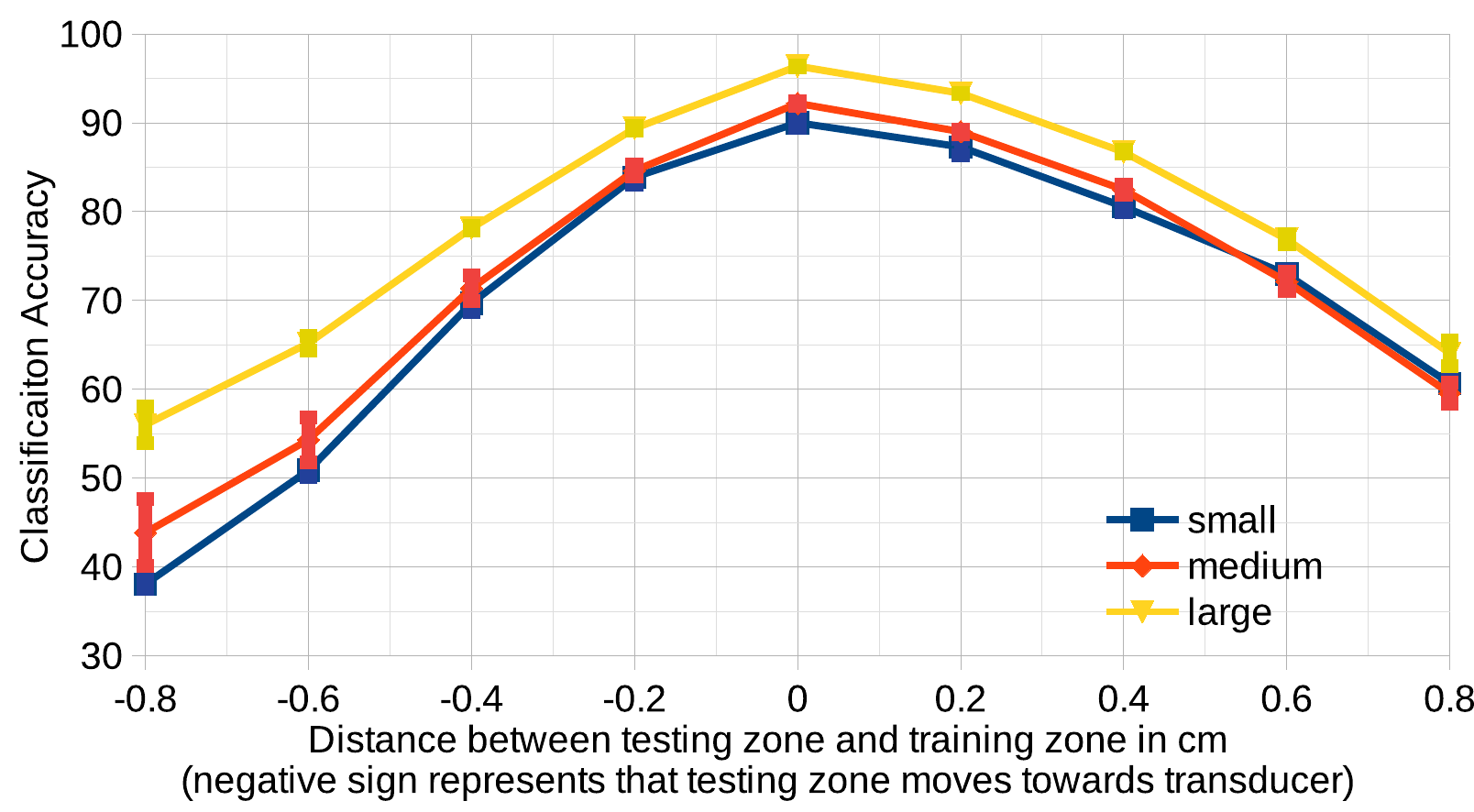}}
\end{tabular}
\caption{Sweeping testing zone center for a network trained by using the on focus zone: Classification Accuracy vs distance between the training and testing zones for different data-set sizes. The colors indicate the size of the training set. The blue color is for 675 patches which is labeled as small, the red color is for 2,700 patches which is labeled as medium and the yellow is for 13,500 patches which is labeled as large.}
\label{fig:datasize_distance}
\endgroup
\end{figure}

In Fig. \ref{fig:datasize_distance}, classification accuracy is plotted as the testing zone center is swept by 0.8 cm towards and away from the transducer around the training zone center. We trained a neural network by using patches from the on focus zone centered at 2 cm depth, and then we tested the neural network with patches from zones centered at 1.2 cm, 1.4 cm, 1.6 cm, 1.8 cm, 2 cm, 2.2 cm, 2.4 cm, 2.6 cm and 2.8 cm, respectively. Overall, the y axis represents classification accuracy and the x axis represents the relative distance between the testing zone and the training zone. For instance, a value of -0.8 means that the testing zone is 0.8 cm closer to transducer than the training zone and +0.8 means that the testing zone is 0.8 cm farther away from transducer than the training zone. We repeat the experiments for different sizes of training sets. We used 675 image patches, 2,700 image patches and 13,500 image patches in the training which correspond to 25 ultrasound images, 100 ultrasound images and 500 ultrasound images, respectively. In the figure, colors indicate the size of the training set. Epoch numbers and learning rates in the training were chosen as 2,000 and 5e-6 for 25 ultrasound images, 1,500 and 1e-5 for 100 ultrasound images, 400 and 1e-5 for 500 ultrasound images.

Several observations can be made form Fig. \ref{fig:datasize_distance}. For the small and medium sets, when the testing zone moved closer to the transducer by 0.4 cm, classification accuracy dropped to 70 percent. However, when the testing zone moved away from the transducer by 0.4 cm, classification accuracy remained above 80 percent. Similarly, for the large set, when the testing zone moved closer to the transducer by 0.4 cm, classification accuracy dropped to below 80 percent. However, when the testing zone moved away from the transducer by 0.4 cm, classification accuracy remained well above 85 percent. Similar observations can be made at other spatial locations as well.
%For the small data-set (blue color), error bar heights are 0.12, 0.27, 0.93, 2.43, 1.00, 1.15, 2.63, 4.72, 5.92. For the medium data-set (orange color), error bar heights are 0.00, 0.82, 1.37, 2.38, 0.39, 0.56, 4.9, 11.45, 10.49 and for the large data-set (yellow color), error bar heights are 0.00, 0.54, 1.97, 3.22, 0.08, 0.74, 3.95, 5.65, 2.31. Error bar heights listed above are given in an order from -1 cm to 1 cm in 0.25 cm increments.  

\begin{figure}[hbt!]
\begingroup
    \centering
    \begin{tabular}{c}
\hspace{-0.3cm}{ \includegraphics[width=1\linewidth]{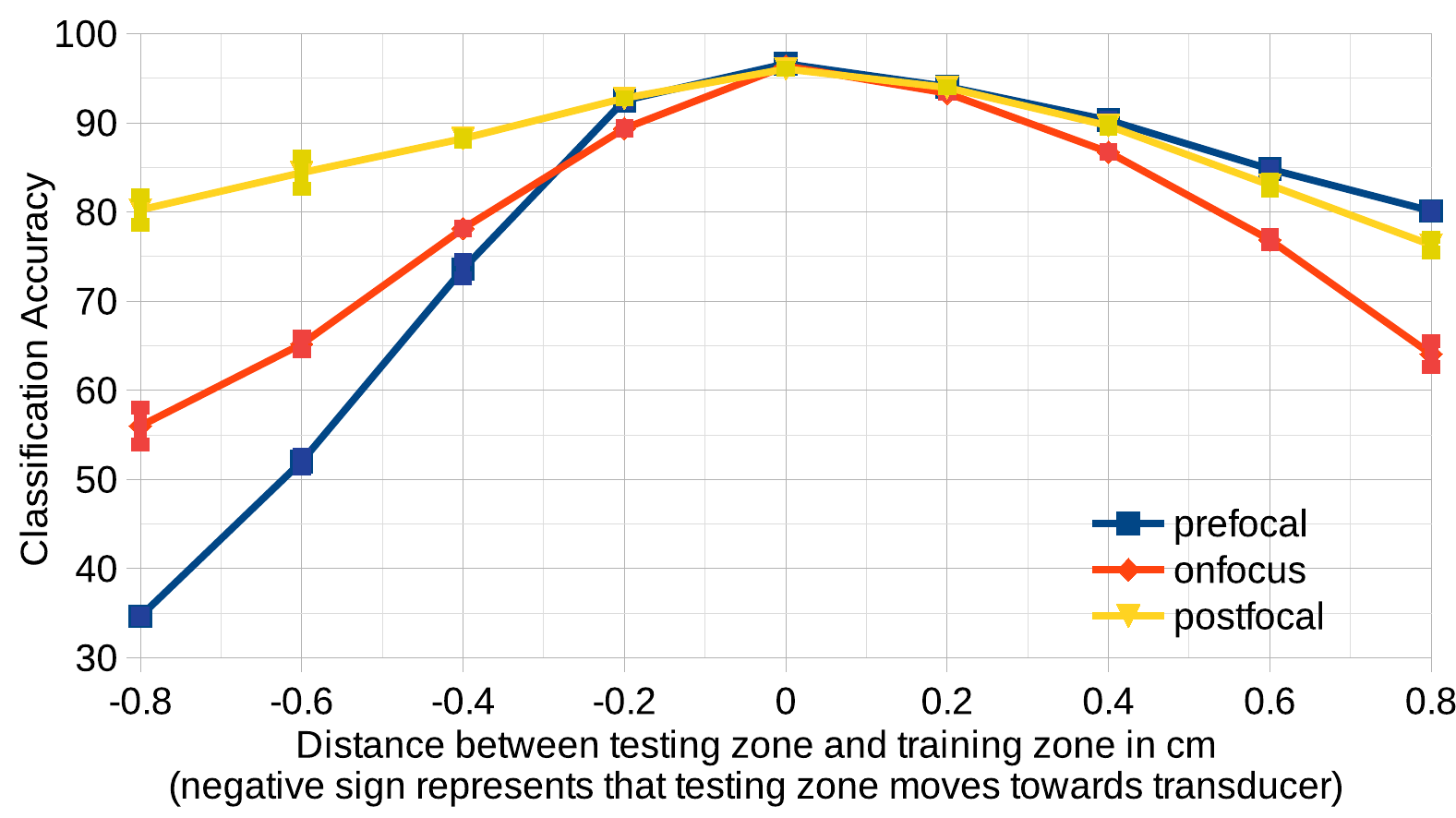}}
\end{tabular}
\caption{Sweeping testing zone center for networks trained on different zones: Classification Accuracy vs distance between training and testing zones. Colors represent the training zone. The blue color is for the pre-focal zone, which is labeled as prefocal. The orange color is for the on focus zone, which is labeled as onfocus. The yellow color is for the post focal zone, which is labeled as postfocal.}
\label{fig:zones_distance}
\endgroup
\end{figure}

In Fig. \ref{fig:zones_distance}, similar to Fig. \ref{fig:datasize_distance}, we plot classification accuracy as the y axis and relative distance between testing zone and training zone as the x axis. In this figure, we experiment with the pre-focal zone and the post focal zone in addition to the on focus zone. When we trained AlexNet by using patches from the pre-focal zone, we tested the network with patches centered at 0.6 cm, 0.8 cm, 1 cm, 1.2 cm, 1.4 cm, 1.6 cm, 1.8 cm, 2 cm and 2.2 cm. When we trained a CNN by using patches from the post focal zone, we tested the network with patches centered at 1.8 cm, 2 cm, 2.2 cm, 2.4 cm, 2.6 cm, 2.8 cm, 3 cm, 3.2 cm and 3.4 cm. In this result, we used fixed training set size, which was 13,500 image patches or 500 ultrasound images. In the figure, colors represent the training zone. Epoch numbers and learning rates in the training were chosen as 400 and 1e-5 for all zones.
%For pre-focus zone, error bar heights are 0.00, 0.00, 2.84, 3.99, 0.21, 1.95, 0.96, 0.62, 2.47. For post focus zone, error bar heights are 0.00, 1.25, 0.98, 0.41, 0.00, 0.15, 0.26, 0.38, 0.71. Error bar heights for on focus zone are given in the previous paragraph.  

Several observations can be made from Fig. \ref{fig:zones_distance} results. When the testing zone was closer to the transducer by 0.4 cm, classification accuracies were slightly lower than 90 percent, slightly lower than 80 percent and 75 percent for the post focal zone, the on focus zone and the pre-focal zone, respectively. However, when the testing zone moved away from the transducer by 0.4 cm, classification accuracies were around 90 percent for all zones. Second, we observed that the post focal zone was the most robust zone against the shift in the testing. Classification accuracy for the post focal zone remained approximately above 80 percent in all shifts.

\begin{figure}[hbt!]
\begingroup
    \centering
    \begin{tabular}{c}
\hspace{-0.3cm}{ \includegraphics[width=1\linewidth]{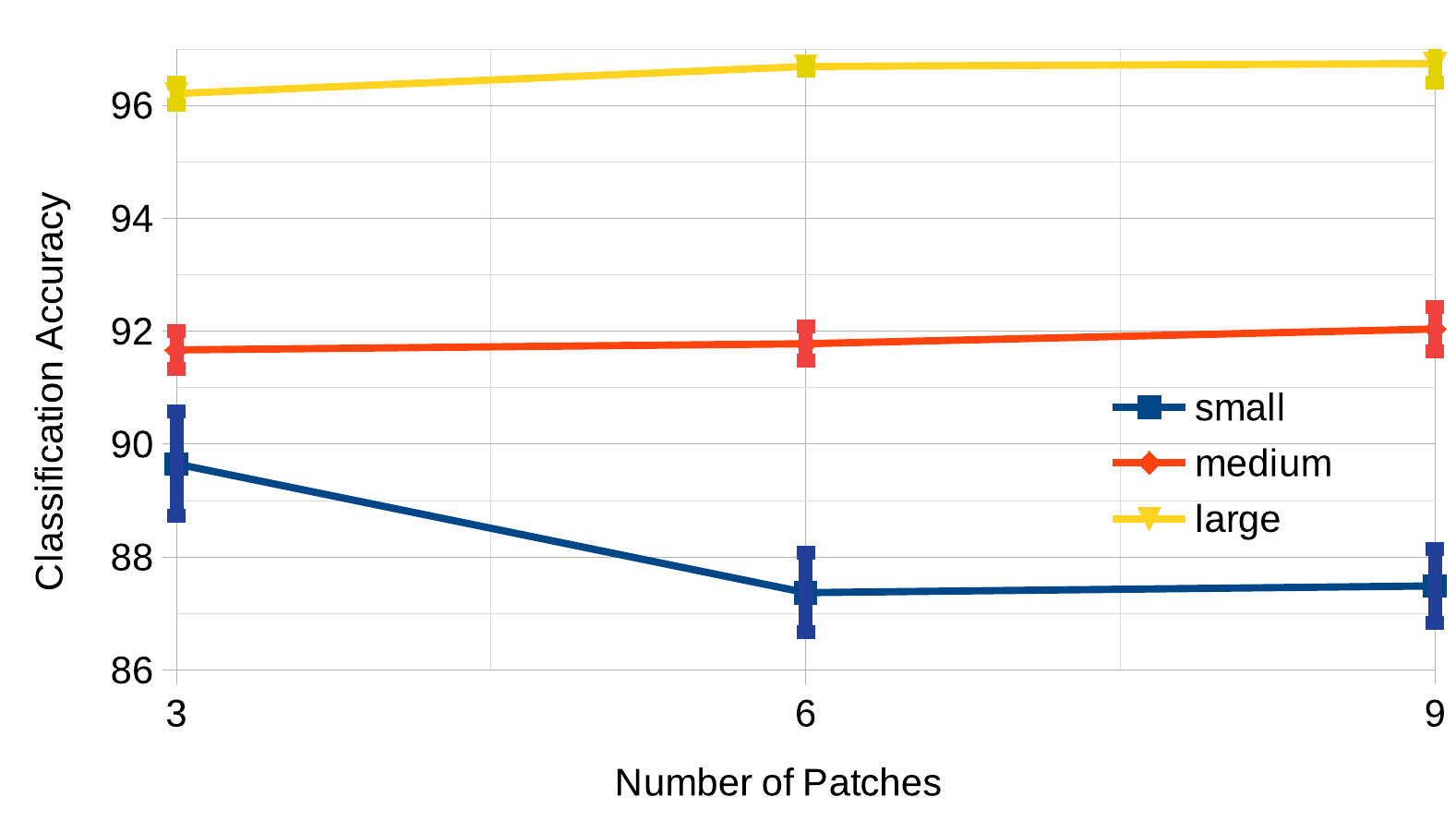}}
\end{tabular}
\caption{Classification accuracy vs axial zone width for the on focus zone. The colors indicate the size of the training set. The blue color is for 25 ultrasound images which is labeled as small, the red color is for 100 ultrasound images which is labeled as medium and the yellow is for 500 ultrasound images which is labeled as large.}
\label{fig:zone_width}
\endgroup
\end{figure}

In Fig. \ref{fig:zone_width}, we plot classification accuracy as the y axis and axial zone width as the x axis for the on focus zone. In \textit{Zone Training}, we extract three overlapping patches per ultrasound image as described in Section \ref{section:dataset} and shown in Fig. \ref{fig:horizontal_patch_extract}. In this result, we make an exception to experiment with zone width, which is defined in terms of number of patches. We now extract 3, 6 and 9 overlapping patches from each ultrasound image for the on focus zone and these numbers form the x axis. Specifically, extracting 3 patches coincides with the original on focus zone definition, while extracting 6 patches coincides with merging pre-focal and on focus zones; and extracting 9 patches coincides with \textit{Regular Training}. Additionally, we used three different sizes for the training set. We used 25 ultrasound images, which corresponds to 675, 1,350, 2,025 training image patches when we extract 3, 6 and 9 patches from each ultrasound image, respectively. Similarly, we used 100 ultrasound images, which corresponds to 2,700, 5,400, 8,100 training image patches and we used 500 ultrasound images, which corresponds to 13,500, 27,000, 40,500 training image patches. As a side note, for this graph, we used the same training and testing zones, unlike the previous two graphs, and colors in the graph represent training set sizes. Moreover, epoch numbers and learning rates were were chosen in accordance with the previous figures. From the Fig. \ref{fig:zone_width}, one can observe that for the small data set size increasing the number of patches, i.e., broadening the zone size, resulted in poorer classification.

\begin{figure}[hbt!]
\begingroup
    \centering
    \begin{tabular}{c}
\hspace{-0.3cm}{ \includegraphics[width=1\linewidth]{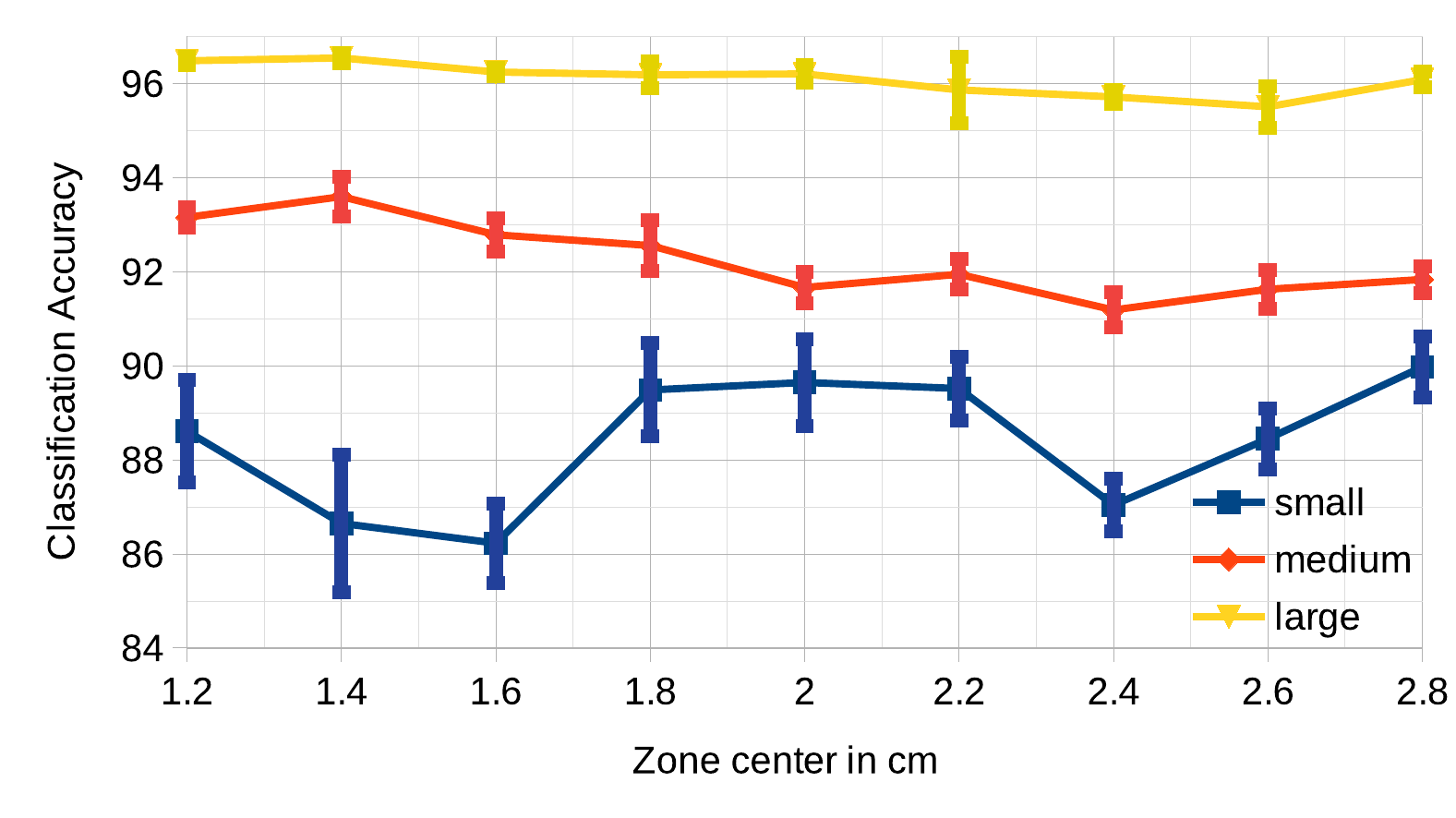}}
\end{tabular}
\caption{Classification accuracy vs zone center for different data-set sizes. Colors represent training set sizes where the blue color is for 675 patches which is labeled as small, the red color is for 2,700 patches which is labeled as medium and the yellow is for 13,500 patches which is labeled as large.}
\label{fig:acc_distance}
\endgroup
\end{figure}

In Fig. \ref{fig:acc_distance}, we plot classification accuracy as the y axis and zone center as the x axis. For this graph, we tested and trained networks from the same zone while sweeping the zone center axially. We trained and tested our networks for zones  centered at 1.2 cm, 1.4 cm, 1.6 cm, 1.8 cm, 2.0 cm, 2.2 cm, 2.4 cm, 2.6 cm and 2.8 cm. We repeat the experiments for different sizes of training sets. We used 675 image patches (25 ultrasound images), 2,700 image patches (100 ultrasound images) and 13,500 image patches (500 ultrasound images) in the training. Epoch numbers and learning rates were chosen in accordance with the previous figures.

% {\color{red}
% -Additional Experiment Idea for Zone Width: Acc. vs Zone Width for different zones
% }

\subsection{Training Set Size vs Classification Accuracy}
\label{sec:data_size}

% \begin{figure}[hbt!]
% \begingroup
%     \centering
%     \begin{tabular}{c}
% \hspace{-0.3cm}{ \includegraphics[width=1\linewidth]{pictures/accvssize_alexnet_new2.png}}
% \end{tabular}
% \caption{Classification accuracy vs training set size in terms of images for the horizontal patch extraction. The brown color represents \textit{Regular Training}, The green color represents \textit{Depth Aware Training}, the orange color represents the on focus zone, the light blue color represents the pre-focal zone centered at 1 cm, dark blue color represents the pre-focal zone centered at 1.5 cm, the yellow represents the post focal zone centered at 2.5 cm and the dark green represents the post focal zone centered at 3 cm. In the experiments, one standard deviation ranged from 0.04 to 5.72. They weren't shown as vertical bars to keep the figure interpretabil.}
% \label{fig:acc_size}
% \endgroup
% \end{figure}

This section compares the performance of \textit{Zone Training} against that of \textit{Regular Training} and \textit{Depth-Aware Training} under various data conditions from low data size regimes to larger data size regimes to investigate if \textit{Zone Training} is more successful when there is a low amount of data. Tables \ref{table:acc0} - \ref{table:acc5} are confusion matrices that list the classification accuracies for different training and testing strategies by using training set sizes of 10, 25, 50, 100, 200 and 500 ultrasound images. Rows represent training strategies: The first row, denoted as pre-focal, is for training with patches from the pre-focal zone. The second row, denoted as on focus, is for training with patches from the on focus zone. The third row, denoted as post focal, is for training with patches from the post focal zone. The fourth row, denoted as regular, is for training with \textit{Regular Training} strategy. The last row, denoted as depth-aware, is for training with \textit{Depth-Aware Training} strategy. Columns represent testing strategies: testing with patches from the pre-focal zone, testing with patches from the on focus zone, testing with patches from the post focal zone and testing with complete field of view, respectively, from first to last column. Epoch numbers and learning rates were were chosen in accordance with the previous figures, which were 2,500 and 5e-6 for 10 ultrasound images, 2,000 and 5e-6 for 25 ultrasound images, 2,000 and 1e-5 for 50 ultrasound images, 1,500 and 1e-5 for 100 ultrasound images, 1,000 and 1e-5 for 200 ultrasound images, 400 and 1e-5 for 500 ultrasound images.

The tables verify that \textit{Zone Training} had better classification accuracy than \textit{Regular Training} and \textit{Depth-Aware Training} in the low data regime. When we used 10, 25 or 50 ultrasound images in training, \textit{Zone Training} performed 1-5 percent better than \textit{Regular Training} and 1-4 percent better than \textit{Depth-Aware Training}. Additionally, \textit{Depth-Aware Training} performed approximately 1 percent better than \textit{Regular Training} for all training set sizes. Lastly, the performance of the zones varied as the size of the training set was reduced. The classification accuracy dropped around 17 percent when we used 10 ultrasound images in training in comparison to 500 ultrasound images for the pre-focal pattern. For the on focus pattern, the same percentage drop was approximately 12 points and for the post focal pattern, the same percentage drop was around 10 points.
% \begin{table}[htb!]
%     \centering
% \caption{Classification Accuracies with 100 ultrasound images }
% {\small
% \begin{tabular}{|p{1cm}|p{1.4cm}|p{1.4cm}|p{1.4cm}|p{1.4cm}|}
%  \hline
% & Pre & On& Post & Regular \\
% \hline
% Pre & 92.9$\pm$4.1  & 75.2$\pm$4.9 & 86.1$\pm$3.7 & 64.9$\pm$3.0 \\
% \hline
% On& 48.9$\pm$3.4 & 95.3$\pm$3.7 & 94.8$\pm$5.1 & 64.3$\pm$1.6 \\
% \hline
% Post & 33.1$\pm$0.6 & 76.3$\pm$4.8 & 98.3$\pm$0.8 & 68.7$\pm$1.2\\
% \hline
% Regular & 61.0$\pm$3.2 & 91.2$\pm$3.9 & 96.5$\pm$1.3 &88.9$\pm$2.0\\
% \hline
% \end{tabular}
% }
% \label{table:acc1}
% \end{table}

\begin{table}[htb!]
    \centering
\caption{Classification Accuracies with 10 ultrasound images}
{\small
\begin{tabular}{|p{2cm}|p{1.4cm}|p{1.4cm}|p{1.4cm}|}
 \hline
& Pre & On& Post \\
\hline
Pre-Focal & 81.01$\pm$3.20  &75.18$\pm$3.67 &  38.99$\pm$1.91\\
\hline
On Focus& 53.00$\pm$4.25 & 86.77$\pm$2.81 & 74.56$\pm$3.33 \\
\hline
Post Focal & 42.78$\pm$3.90 & 74.00$\pm$4.82 & 86.96$\pm$1.88 \\
\hline
Regular & 78.62$\pm$2.53 & 83.52$\pm$2.00 & 85.14$\pm$2.13\\
\hline
Depth-Aware & 80.74$\pm$2.43 & 83.96$\pm$3.17 & 87.27$\pm$2.45\\
\hline
\end{tabular}
}
\label{table:acc0}
\end{table}

\begin{table}[htb!]
    \centering
\caption{Classification Accuracies with 25 ultrasound images}
{\small
\begin{tabular}{|p{2cm}|p{1.4cm}|p{1.4cm}|p{1.4cm}|}
 \hline
& Pre & On& Post \\
\hline
Pre-Focal & 86.65$\pm$2.91  & 77.70$\pm$4.14 & 42.73$\pm$9.93 \\
\hline
On Focus& 54.19$\pm$2.86 & 89.65$\pm$1.85 & 73.07$\pm$3.11 \\
\hline
Post Focal & 43.11$\pm$3.61 & 73.17$\pm$4.40 & 88.45$\pm$1.29 \\
\hline
Regular & 81.64$\pm$2.80 & 87.49$\pm$1.31 & 87.35$\pm$1.73\\
\hline
Depth-Aware &82.79$\pm$1.52 & 87.89$\pm$0.91 & 88.81$\pm$1.23\\
\hline
\end{tabular}
}
\label{table:acc1}
\end{table}

% \begin{table}[htb!]
%     \centering
% \caption{Classification Accuracies with 200 ultrasound images }
% {\small
% \begin{tabular}{|p{1cm}|p{1.4cm}|p{1.4cm}|p{1.4cm}|p{1.4cm}|}
%  \hline
% & Pre & On & Post & Regular \\
% \hline
% Pre & 96.5$\pm$2.3& 72.7$\pm$2.8& 81.3$\pm$5.8 & 65.4$\pm$1.4\\
% \hline
% On& 49.4$\pm$2.6 & 96.5$\pm$1.6 & 95.4$\pm$2.4 & 60.9$\pm$1.6\\
% \hline
% Post & 33.4$\pm$0.3 & 80.6$\pm$3.7 & 99.2$\pm$0.9 & 66.6$\pm$1.5\\
% \hline
% Regular & 52.5$\pm$4.1 & 95.4$\pm$2.1 & 98.3$\pm$1.4 & 96.2$\pm$0.8\\
% \hline
% \end{tabular}
% }
% \label{table:acc2}
% \end{table}

\begin{table}[htb!]
    \centering
\caption{Classification Accuracies with 50 ultrasound images}
{\small
\begin{tabular}{|p{2cm}|p{1.4cm}|p{1.4cm}|p{1.4cm}|}
 \hline
& Pre & On& Post \\
\hline
Pre-Focal & 91.66$\pm$0.93  & 83.91$\pm$2.41 & 56.60$\pm$8.91 \\
\hline
On Focus& 53.42$\pm$3.12 & 90.42$\pm$0.96 & 73.98$\pm$3.51 \\
\hline
Post Focal & 44.75$\pm$3.10 & 77.87$\pm$2.98 & 89.63$\pm$1.12 \\
\hline
Regular & 89.66$\pm$1.61 & 89.23$\pm$1.85 & 89.01$\pm$1.06\\
\hline
Depth-Aware & 90.15$\pm$1.90 & 90.32$\pm$1.49 & 89.91$\pm$1.95\\
\hline
\end{tabular}
}
\label{table:acc2}
\end{table}

% \begin{table}[htb!]
%     \centering
% \caption{Classification Accuracies with 300 ultrasound images }
% {\small
% \begin{tabular}{|p{1cm}|p{1.4cm}|p{1.4cm}|p{1.4cm}|p{1.4cm}|}
%  \hline
% & Pre & On & Post & Regular \\
% \hline
% Pre & 97.3$\pm$3.3& 75.2$\pm$2.0& 79.7$\pm$5.1 & 64.8$\pm$1.9\\
% \hline
% On& 46.4$\pm$2.0 & 97.9$\pm$1.4 & 93.9$\pm$2.4 & 59.2$\pm$1.9\\
% \hline
% Post & 33.5$\pm$0.3 & 74.2$\pm$2.8 & 99.5$\pm$0.5 & 68.1$\pm$0.9\\
% \hline
% Regular & 49.4$\pm$3.1 & 95.0$\pm$2.4 & 97.7$\pm$2.2 & 96.9$\pm$1.3\\
% \hline
% \end{tabular}
% }
% \label{table:acc3}
% \end{table}

\begin{table}[htb!]
    \centering
\caption{Classification Accuracies with 100 ultrasound images}
{\small
\begin{tabular}{|p{2cm}|p{1.4cm}|p{1.4cm}|p{1.4cm}|}
 \hline
& Pre & On& Post \\
\hline
Pre-Focal & 93.60$\pm$0.85  & 84.09$\pm$1.51 & 62.19$\pm$8.20 \\
\hline
On Focus& 57.60$\pm$3.94 & 91.67$\pm$0.67 & 74.40$\pm$2.06 \\
\hline
Post Focal & 48.62$\pm$3.13 & 78.53$\pm$4.54 & 91.63$\pm$0.82 \\
\hline
Regular & 93.36$\pm$1.22 & 92.04$\pm$0.79 & 92.08$\pm$0.57\\
\hline
Depth-Aware & 93.58$\pm$1.17 & 92.61$\pm$0.99 & 92.77$\pm$1.06\\
\hline
\end{tabular}
}
\label{table:acc3}
\end{table}

\begin{table}[htb!]
    \centering
\caption{Classification Accuracies with 200 ultrasound images}
{\small
\begin{tabular}{|p{2cm}|p{1.4cm}|p{1.4cm}|p{1.4cm}|}
 \hline
& Pre & On& Post \\
\hline
Pre-Focal & 94.69$\pm$0.41  & 84.43$\pm$0.56 & 63.77$\pm$3.93 \\
\hline
On Focus& 61.45$\pm$4.47 & 94.19$\pm$0.35 & 76.06$\pm$2.28 \\
\hline
Post Focal & 57.95$\pm$5.66 & 83.79$\pm$1.98 & 93.41$\pm$0.53 \\
\hline
Regular & 94.79$\pm$0.35 & 94.50$\pm$0.41 & 94.00$\pm$0.35\\
\hline
Depth-Aware & 95.17$\pm$0.27 & 95.00$\pm$0.29 &95.08$\pm$0.36\\
\hline
\end{tabular}
}
\label{table:acc4}
\end{table}

\begin{table}[htb!]
    \centering
\caption{Classification Accuracies with 500 ultrasound images}
{\small
\begin{tabular}{|p{2cm}|p{1.4cm}|p{1.4cm}|p{1.4cm}|}
 \hline
& Pre & On& Post \\
\hline
Pre-Focal & 96.55$\pm$0.22  & 84.53$\pm$0.88 & 66.53$\pm$5.62 \\
\hline
On Focus& 62.25$\pm$4.92 & 96.21$\pm$0.36 & 75.87$\pm$3.05 \\
\hline
Post Focal & 58.97$\pm$8.04 & 82.34$\pm$2.74 & 95.51$\pm$0.89 \\
\hline
Regular & 97.09$\pm$0.77 & 96.74$\pm$0.67 & 96.75$\pm$0.32\\
\hline
Depth-Aware & 97.40$\pm$0.10 & 97.32$\pm$0.16 & 97.15$\pm$0.16\\
\hline
\end{tabular}
}
\label{table:acc5}
\end{table}
\section{Discussion}
\label{sec:discussion}

We proposed a DL training strategy, named \textit{Zone Training}, where we split the complete field of view into zones such as the pre-focal, the on focus and the post focal zones. Then, we trained separate networks for each zone. We investigated \textit{Zone Training} thoroughly by experimenting with zone definitions and their behavior under different training set sizes.

The figures provide several important observations. From Fig. \ref{fig:datasize_distance}, we observed that as the testing zone moved towards the transducer, classification accuracy dropped faster and it was valid for small, medium and large training set sizes. The observation indicates that the pre-focal diffraction pattern was more complicated and it changed faster than the post focal diffraction pattern. 

In Fig. \ref{fig:zones_distance}, we quantified how classification accuracy decreased when the testing zone moved away from the training zone for training with the on focus zone, the pre-focal zone and the post focal zone. First, we observed that when the testing zone was closer to the transducer, classification accuracies dropped faster for the pre-focal and on focus zones. For the post focal zone, classification accuracies were relatively symmetric around the zone center. This further verified our previous observation stating that the pre-focal pattern was more complicated and changed quickly in comparison to the post focal pattern. Another observation was that for the pre-focal training, classification accuracy deteriorated slowly when the testing zone moved away from the transducer in comparison to the testing zone moving towards the transducer, which further illustrates the complicated behaviour of the pre-focal pattern.

In Fig. \ref{fig:zone_width}, we investigated the relationship between classification accuracy and zone width in terms of overlapping patches for the on focus zone. First, we observed that as we increased the number of patches, classification accuracy remained relatively constant for the medium size training set, while classification accuracy slightly increased for the large size training set. However, classification accuracy dropped as we increased the number of patches for the small size training set. Specifically, classification accuracy dropped to around 87 percent from 90 percent as we increased the number of patches from 3 to 6 and it stayed relatively constant when we increased the number of patches to 9. These observations indicate that \textit{Zone Training} was more robust when the training set size was smaller. However, \textit{Regular Training} can be preferable when the training set size was larger.

In Fig. \ref{fig:acc_distance}, we determined the best zone location axially in terms of classification accuracy. We observed that the relationship between classification accuracy and the zone location depended on the training set size. For the large training set size, the classification accuracy stayed relatively constant around 96 percent in all axial locations. For the medium training set size, the classification accuracy degraded from approximately 94 percent to 92 percent when the zone location moved from 1.2 cm to 2.8 cm. That indicates the pre-focal zones are the most desirable zones for the medium size training set size. However, for the small training set size, the zones around the on focus zone are the most desirable. Fig. \ref{fig:acc_distance} is useful to determine the most optimal zone center to characterize tissue samples for different training conditions. However, using a single zone is only meaningful when the phantoms are uniform and we don't loose any information by discarding other zones in our decision process. If there is some spatial information to be taken advantage of in our classification decision or we want to increase classification accuracies by using all information that we have, then we need to separate the complete field of view into multiple zones and train multiple expert networks to be used in a voting schema. In that case, Fig. \ref{fig:acc_distance} is still useful for determining which expert network should have higher effect in a voting schema.

%First, we observed that \textit{Regular Training} had the lowest classification accuracy for all training set sizes of 100, 200, 300, 400 and 500 ultrasound images. On the other hand, the post focal zones had the highest accuracy, which was very close to 100 percent for all training set sizes. Among all zones, the pre-focal at 1.5 cm had the lowest classification accuracy, and therefore, it required more training data. The most remarkable observation from this figure was that \textit{Regular Training} required at least 500 ultrasound images to achieve similar performance with the post focal training by using only 100 ultrasound images. In addition, \textit{Regular Training} required at least 200 images to achieve similar performance with the on focus training and the pre-focal training by using only 100 ultrasound images. These observations verify that in order for \textit{"Regular Training"} to achieve similar accuracies as \textit{"Zone Training"}, it requires more training data. 

In Tables \ref{table:acc0} - \ref{table:acc5}, we presented confusion matrices to quantify classification accuracies with respect to different training set sizes, which are 10, 25, 20, 100, 200 and 500 ultrasound images. First, \textit{Zone Training} had better classification accuracy than \textit{Regular Training} when the training data was scarce. However, when the training data size was larger, \textit{Regular Training} performed better than \textit{Zone Training}. For example, when we used 200 or 500 ultrasound images in training, \textit{Regular Training} performed around 1 percent better than \textit{Zone Training} for all zones. However, when we used 10, 25 or 50 ultrasound images in training, \textit{Zone Training} performed better than \textit{Regular Training}. When we used 100 ultrasound images, \textit{Regular Training} and \textit{Zone Training} performed similarly. Second, \textit{Depth-Aware Training} was always better than \textit{Regular Training}. Third, \textit{Zone Training} was slightly better than \textit{Depth-Aware Training} when the training data was scarce. When we used 10, 25 or 50 ultrasound images in training, \textit{Zone Training} performed better than \textit{Depth-Aware Training} for the pre-focal and on focus zones. However, they performed similarly for the post focal zone. Lastly, the post-focal pattern was more robust against decreasing training set size in comparison to the on focus pattern; and the on focus pattern was more robust in comparison to the pre-focal pattern for all training strategies (\textit{Zone, Regular and Depth-Aware Training}).

In this work, our proposed method was applied to tissue classification. However, it can be applicable to other DL applications such as detection, segmentation and image formation. Therefore, \textit{Zone Training} can be tested for different applications. Additionally, it would be interesting to test \textit{Zone Training} with different types of neural network structures even though \textit{Zone Training} is not directly related to the neural network structure. Moreover, the optimal number of zones should be investigated in greater detail. The optimal number of zones can change from problem to problem depending on the imaging substrates, imaging system, problem complexity and imaging settings. 
It is also important to consider varying zone widths for different focal zones, as this can have a significant impact on the optimal number of zones, e.g., breaking the pre-focal zone into multiple, smaller zones might improve the accuracy in the pre-focal region. Furthermore, different patch sizes including pixel-wise classification, which is known as image segmentation, can be investigated within the context of \textit{Zone Training}. Ultimately, we would like to identify the most data efficient DL algorithm in the context of tissue characterization and \textit{Zone Training} can be a useful tool in low data regimes for ultrasound imaging.

\section{Conclusion}
\label{sec:conclusion}
We have presented a data-efficient DL training strategy for tissue classification with biomedical ultrasound imaging, which we named \textit{Zone Training}. \textit{Zone Training} has the ability to maintain high classification accuracy, while reducing training set size. Therefore, it should be considered as a robust approach for DL powered ultrasound imaging in the context of tissue characterization.
% \section*{Acknowledgment}
\bibliographystyle{IEEE_ECE}
% Put references in BibTeX format in thesisrefs.bib.
\bibliography{paper}
%\todo[inline]{The following applies to about 10 of the references: "IEEE publications must list names of all authors, up to \emph{six} names. If there are more than six names listed, use the primary author’s name followed by et al." (See https://ieeeauthorcenter.ieee.org/wp-content/uploads/IEEE-Reference-Guide.pdf) --Done.}
%\todo[inline]{Use the standard abbreviations for publication names, e.g., 
%"IEEE Transactions on Ultrasonics, Ferroelectrics, and Frequency Control" =
%"IEEE Trans. Ultras. Ferroel. Freq. Cont.", etc. You can get an extra bib file with all the standard abbreviations and include it. For some sources, you can use extreme abbreviations (edit your bib file manually), because we are submitting to an IEEE journal and all readers would know what "ICASSP 2017" is:
%"2017 IEEE International  Conference  on  Acoustics,  Speech  and  Signal  Processing(ICASSP).    IEEE, 2017" =
%"ICASSP 2017"--Done}
% \include{supp}
\end{document}